\documentclass[
    ,final            % use final for the camera ready runs
  ]
  {aipproc}

\layoutstyle{6x9}

\begin{document}

\title{Effective Field Theory for Cold Atoms}

\author{H.-W. Hammer}{
  address={Institute for Nuclear Theory, University of Washington, 
  Seattle, WA 98195-1550, USA}
}

\begin{abstract}
Effective Field Theory (EFT) provides a powerful framework that exploits 
a separation of scales in physical systems to perform systematically
improvable, model-independent calculations. 
Particularly interesting are few-body systems with 
short-range interactions and large two-body scattering length.
Such systems display remarkable universal features. In systems with more 
than two particles, a three-body force with limit cycle behavior is 
required for consistent renormalization already at leading order. 
We will review this EFT and some of its applications in the physics 
of cold atoms. Recent extensions of this approach to the four-body
system and $N$-boson droplets in two spatial dimensions will also be 
discussed.
\end{abstract}

\maketitle

%%%%%%%%%%%%%%%%%%%%%%%%%%%%%%%%%%%%%%%%%%%%
%% MAINMATTER
%%%%%%%%%%%%%%%%%%%%%%%%%%%%%%%%%%%%%%%%%%%%

\section{INTRODUCTION}
The Effective Field Theory (EFT) approach provides a powerful framework 
that exploits the separation of scales in physical systems 
\cite{Kaplan:1995uv}.
Only low-energy (or long-range) degrees of freedom are included
explicitly, with the rest parametrized in terms of the most general
local contact interactions. This procedure exploits the fact that
a low-energy probe of momentum $k$ cannot resolve structures on 
scales smaller than $1/k$. (Note that $\hbar=1$ in this talk.)
Using renormalization,
the influence of short-distance physics on low-energy observables
is captured in a few low-energy constants.
Thus, the EFT describes universal low-energy physics independent of
detailed assumptions about the short-distance dynamics. 
All physical observables can be described in a controlled expansion in 
powers of $kl$, where $l$ is the characteristic low-energy 
length scale of the system. The size of $l$ depends on the system 
under consideration: for a finite range potential, e.g., it is given 
by the range of the potential.
For the systems discussed here, $l$ is of the order of the effective 
range $r_e$.

We will focus on applications of EFT to few-body systems
with large S-wave scattering length $|a| \gg l$. 
For a generic system, the scattering length
is of the same order of magnitude as the low-energy length scale $l$.
Only a very specific choice of the parameters in the underlying theory 
(a so-called {\it fine tuning}) will generate a large scattering length $a$.
Nevertheless, systems with large scattering length can be found in many
areas of physics. The fine tuning can be accidental or it can be 
controlled by an external parameter. 
Examples for an accidental fine tuning are the S-wave scattering 
of nucleons or of $^4$He atoms. 
For alkali atoms close to a Feshbach resonance,
$a$ can be tuned experimentally by adjusting an external magnetic field.
At very low energies these systems behave similarly and
show universal properties associated with the large scattering length.

We will start with a brief review of the EFT for 
few-body systems with large $a$ and then discuss some of the
universal properties predicted in three and two spatial dimensions.

\section{THREE-BODY SYSTEM WITH LARGE SCATTERING LENGTH}

We consider three spatial dimensions first.
For typical momenta $k\sim 1/|a|$, the EFT expansion is in powers of 
$l/|a|$; in leading order we can set $l=0$.
We start with a two-body system of nonrelativistic bosonic atoms 
with large S-wave scattering length $a$ and mass $m$.
At sufficiently low energies, the most general Lagrangian 
may be written as:
\begin{equation}
{\cal L}  =  \psi^\dagger
             \bigg(i\partial_{t}+\frac{\vec{\nabla}^{2}}{2m}\bigg)\psi
 - \frac{C_0}{2} (\psi^\dagger \psi)^2
 - \frac{D_0}{6} (\psi^\dagger\psi)^3 + \ldots\,,
\label{eq:eftlag}
\end{equation}
where the $C_0$ and $D_0$ are nonderivative two- and three-body interaction
terms, respectively. The strength of the $C_0$ term is determined by
the scattering length $a$, while $D_0$ depends on a three-body parameter
to be introduced below.
The dots represent higher-order derivative terms
which are suppressed at low-energies. For momenta $k$ of the order of
the inverse scattering length $1/|a|$, the problem is nonperturbative 
in $ka$. The exact two-body scattering amplitude can be obtained
analytically by summing the so-called {\it bubble diagrams} with the
$C_0$ interaction term. The $D_0$ term does not contribute to two-body
observables. After renormalization, 
the resulting amplitude reproduces the leading order of the 
well-known effective range expansion for the atom-atom
scattering amplitude:~$f_{AA}(k)=(-1/a -ik)^{-1}\,,$
where the total energy is $E=k^2/m$. 
If $a>0$, $f_{AA}$ has a pole at $k=i/a$ corresponding
to a shallow dimer with binding energy $B_2=1/(ma^2)$. 
Higher-order derivative
interactions are perturbative and give the momentum-dependent terms
in the effective range expansion.

\begin{figure}[htb]
\bigskip
\centerline{\includegraphics*[width=14cm,angle=0]{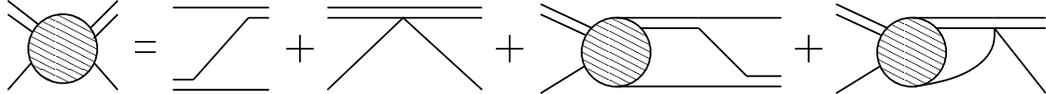}}
\caption{Integral equation for the atom-dimer scattering amplitude.
Single (double) line indicates single atom (two-atom state).}
\label{fig:ineq}
\end{figure}

We now turn to the three-body system. Here, it is useful to introduce
an auxiliary field for the two-atom state (see Ref.~\cite{BHvK99}
for details).
At leading order, the atom-dimer scattering amplitude is given by the 
integral equation shown in  Fig.~\ref{fig:ineq}. 
%%%%%%%%%%%%%%%%%%%%%%%%%%%%%%%%%%%%%%%%%%%%%%%%%%%%%%%%%%%%%%%%%%%%%%%%
%%%%%%%%%%%%%%%%%%%%%%%%%%%%%%%%%%%%%%%%%%%%%%%%%%%%%%%%%%%%%%%%%%%%%%%%
A solid line indicates a single atom and a double line indicates an 
interacting two-atom state (including rescattering corrections).
The integral equation contains contributions from both
the two-body and the three-body interaction terms.
The inhomogeneous term is given by the first two diagrams on the 
right-hand side: the one-atom exchange diagram
and the three-body interaction. The integral equation simply sums
these diagrams to all orders.
After projecting onto S-waves, we obtain the equation
\begin{eqnarray}
{\cal T} (k, p; E) & = & {16 \over 3 a}\, M(k,p;E)
+ {4 \over \pi} \int_0^\Lambda 
{dq \, q^2 \, M(q,p;E)\over  -{1/a} + \sqrt{3q^2/4 -mE
-i \epsilon}}\, {\cal T} (k, q; E)\,,
\label{eq:BHvK}
\end{eqnarray}
for the off-shell atom-dimer scattering amplitude with the inhomogeneous
term 
\begin{eqnarray}
M(k,p;E)&=& {1 \over 2pk} \ln \left({p^2 + pk + k^2 -mE \over
p^2 - pk + k^2 - mE}\right) + {H(\Lambda) \over \Lambda^2}\,.
\end{eqnarray}
The logarithmic term is the S-wave projected one-atom exchange,
while the term proportional to $H(\Lambda)$ comes from the three-body
force. The S-wave atom-dimer scattering amplitude 
$f_{AD}(k)=[k\cot\delta_0-ik]^{-1}$ is given by the 
solution ${\cal T}$ evaluated at the on-shell 
point:~$f_{AD}(k) = {\cal T} (k, k; E)$ where
$mE= 3k^2/4-1/a^2\,$.
The three-body binding energies $B_3$ are given by those values of $E<0$
for which the homogeneous version of Eq.~(\ref{eq:BHvK}) has a
nontrivial solution.

Note that an ultraviolet cutoff $\Lambda$ has been introduced in
(\ref{eq:BHvK}). This cutoff is required to insure that Eq.~(\ref{eq:BHvK})
has a unique solution. All physical observables, however, must be 
independent of $\Lambda$, which determines the behavior of $H$ as a 
function of $\Lambda$ \cite{BHvK99}:
\begin{eqnarray}
H (\Lambda) = {\cos [s_0 \ln (\Lambda/ \Lambda_*) + \arctan s_0]
\over \cos [s_0 \ln (\Lambda/ \Lambda_*) - \arctan s_0]}\,,
\label{H-Lambda}
\end{eqnarray}
where $s_0=1.00624$ is a transcendental number and $\Lambda_*$
is a three-body parameter introduced by dimensional transmutation. 
This parameter cannot be predicted by the EFT and must be 
determined from a three-body observable.
Note also that $H (\Lambda)$ is periodic and runs on
a limit cycle. When $\Lambda$ is increased by a factor of
$\exp(\pi/s_0)\approx 22.7$, $H (\Lambda)$ returns to its original 
value. 

In summary, two parameters are required to specify a 
three-body system at leading order in $l/|a|$:
they may be chosen as the scattering length $a$ (or equivalently
$B_2$ if $a>0$) and the three-body parameter $\Lambda_*$ \cite{BHvK99}. 

\section{UNIVERSAL PROPERTIES IN 3D}

This EFT confirms and extends the universal predictions for
the three-body system first derived by Efimov \cite{Efi71}. 
The best known example is
the {\it Efimov effect}, the accumulation of infinitely many three-body 
bound states at threshold as $a\to\pm\infty$.
Universality also constrains three-body scattering observables.
The atom-dimer scattering length, e.g., can be expressed in terms of $a$
and $\Lambda_*$ as \cite{Efi71,BH02}
\begin{equation}
a_{12}=a\, (1.46-2.15\tan [s_0 \ln(a\Lambda_*) +0.09])\,
(1\,+\,{\cal O}(l/a))\,,\qquad\qquad a>0\,.
\end{equation}
Note that the log-periodic dependence of $H$ on $\Lambda_*$ 
is not an artefact of the renormalization and also shows up in observables 
like $a_{12}$. This dependence could, e.g.,
be tested experimentally with atoms close to a Feshbach resonance
by varying $a$.
Similar expressions can be obtained for other three-body
observables such as scattering phase shifts as well as
three-body recombination and dimer relaxation
rates \cite{BH02,Braaten:2003yc}.

\begin{figure}[htb]
\centerline{\includegraphics*[width=8cm,angle=0]{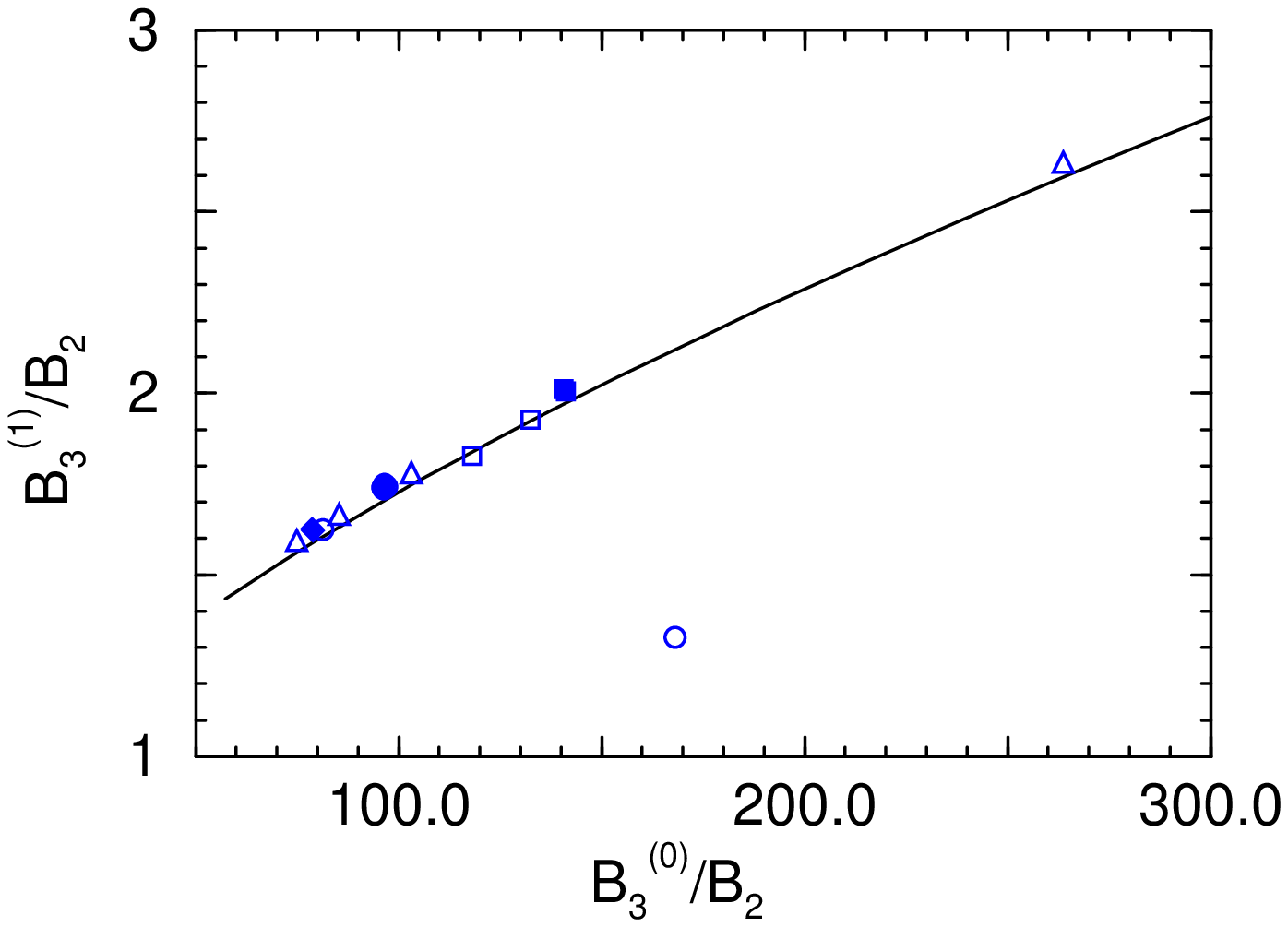}
\quad\includegraphics*[width=6.6cm,angle=0]{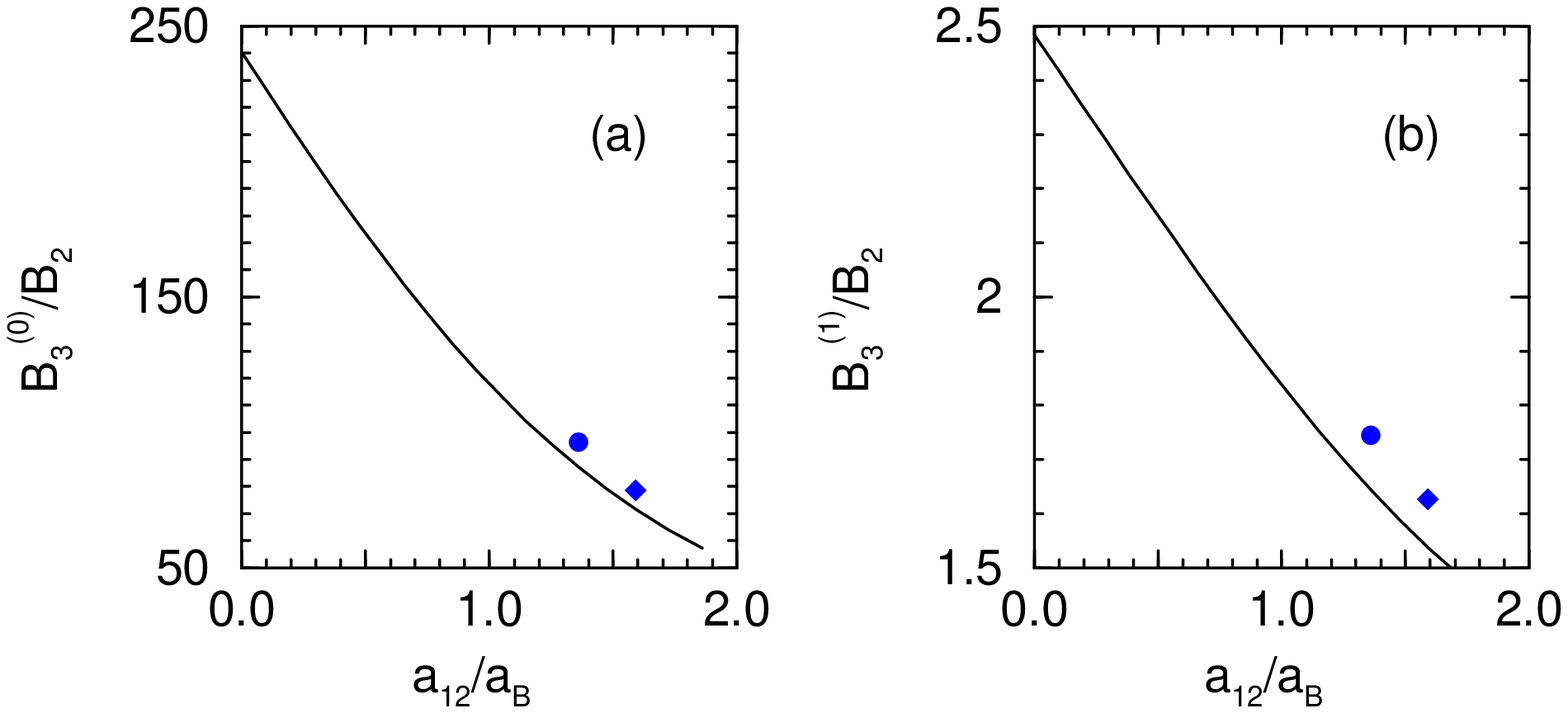}}
\caption{Scaling functions relating the $^4$He trimer ground and excited state
energies (left panel) and the $^4$He trimer ground state and the 
atom-dimer scattering length (right panel). The data points are 
calculations using various methods and $^4$He potentials 
(see Ref.~\cite{MSSK01} and references therein).
}
\label{fig:scale3}
\end{figure}

Since up to corrections of order $l/|a|$,
low-energy three-body observables depend on $a$ and $\Lambda_*$ only, 
they obey non-trivial scaling relations. If dimensionless combinations
of such observables are plotted against each other, the must fall close
to a line parametrized by $\Lambda_*$ \cite{BHvK99,BH02}.
An example of a scaling function relating $^4$He trimer ground and excited 
state energies $B_3^{(0)}$ and $B_3^{(1)}$ is shown in the left panel
of Fig.~\ref{fig:scale3}. 
%%%%%%%%%%%%%%%%%%%%%%%%%%%%%%%%%%%%%%%%%%%%%%%%%%%%%%%%%%%%%%%%%%%%%%%%
%%%%%%%%%%%%%%%%%%%%%%%%%%%%%%%%%%%%%%%%%%%%%%%%%%%%%%%%%%%%%%%%%%%%%%%%
(A related scaling function was obtained in Ref.~\cite{FTD99}.)
The data points show calculations using various approaches and $^4$He 
potentials. Since different potentials have approximately the same
scattering length but include different short-distance physics,
different points on this line correspond to different values of $\Lambda_*$.
The small deviations of the potential model calculations are mainly due to
effective range effects. They are of the order $r_e/a \approx 10\%$
and can be calculated at next-to-leading order in EFT.
The calculation corresponding to the data point far off the universal curve 
can easily be identified as problematic since the deviation from universality
by far exceeds the expected 10\%.  The right panel shows
the scaling function relating the $^4$He trimer ground state 
energy $B_3^{(0)}$  and the atom-dimer scattering length $a_{12}$.
(Note that $a_B\equiv 1/\sqrt{mB_2}$.) A similar scaling relation
is observed in nuclear physics between the spin-doublet neutron-deuteron
scattering length and the triton binding energy and is known
as the Phillips line.

\begin{figure}[htb]
\centerline{\includegraphics*[width=7.2cm,angle=0]{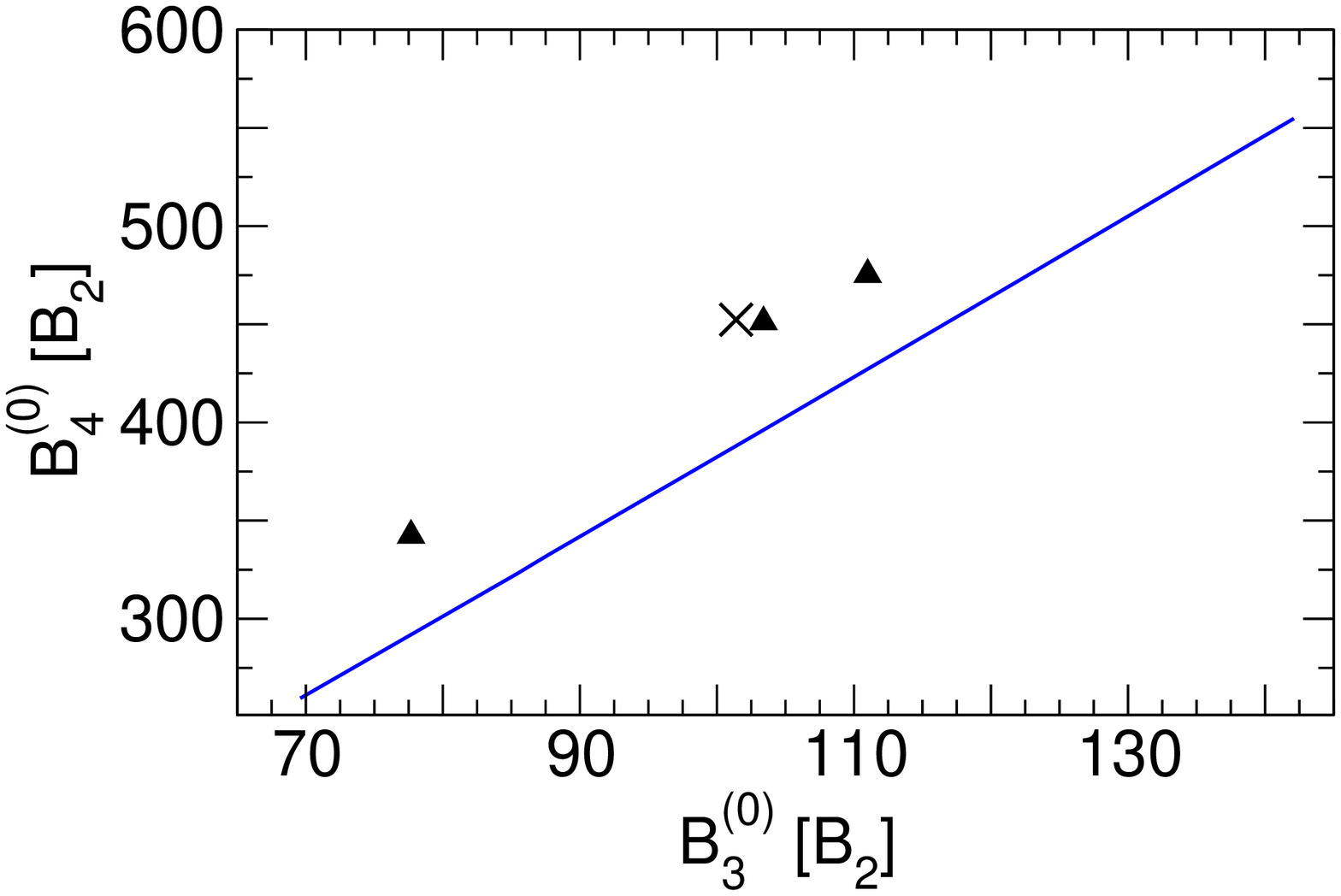}
\quad
\includegraphics*[width=7.2cm,angle=0]{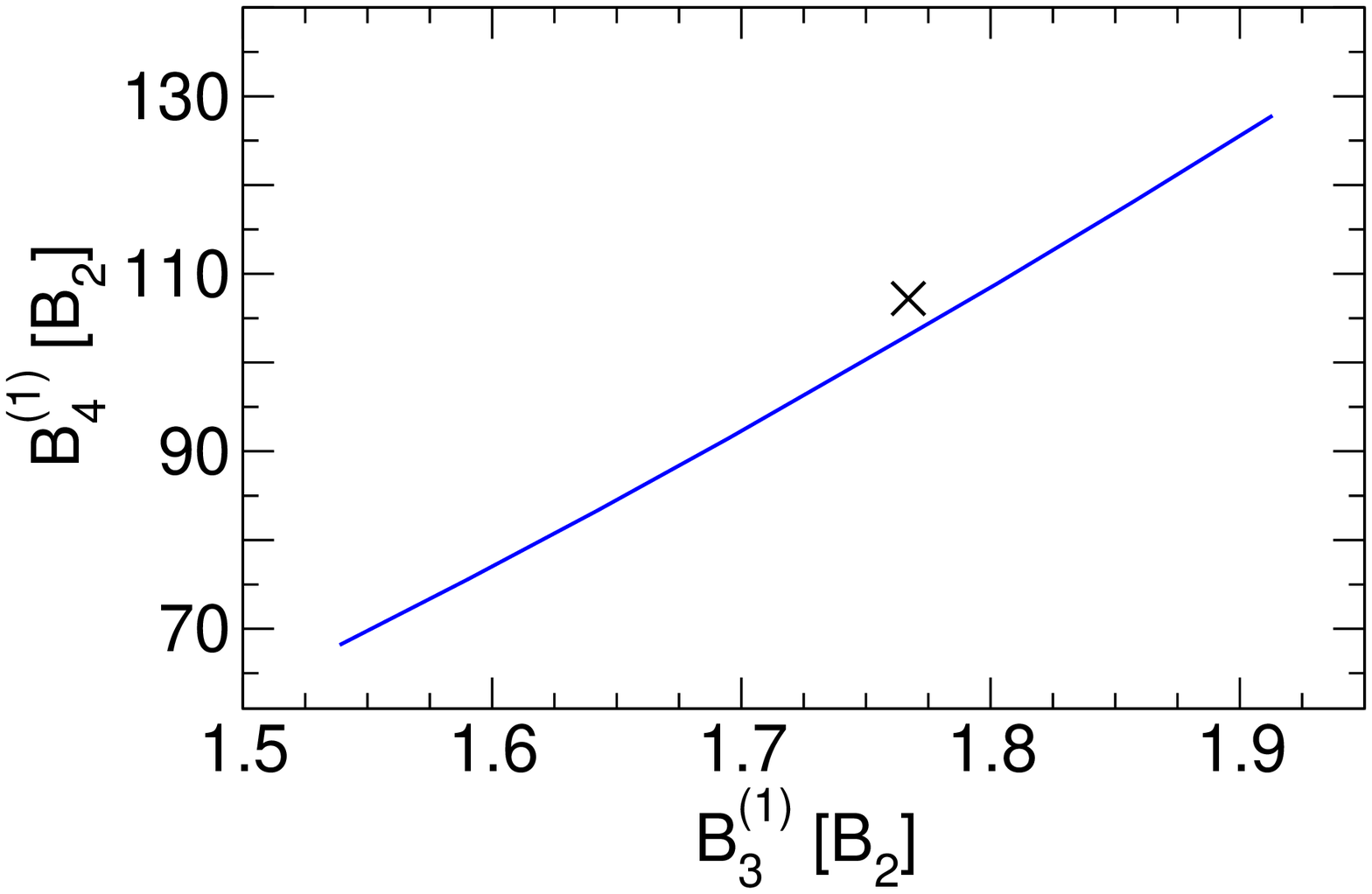}}
\caption{Scaling function relating the $^4$He trimer and tetramer 
ground state (left panel) and excited state (right panel)
energies. The data points are calculations 
using various methods and $^4$He potentials.
}
\label{fig:scale4}
\end{figure}

Recently, we have extended the effective theory for large scattering 
length to the four-body system \cite{Platter:2004qn}. 
It is advantageous in this case to use an effective
quantum mechanics framework, since one can directly start from the 
well-know Yakubovsky equations. We have shown that no four-body
parameter enters at leading order in $l/|a|$. Therefore 
renormalization of the three-body system automatically guarantees 
renormalization of the four-body system. As a consequence, there
are universal scaling functions relating three- and four-body
observables as well. As an example, we show the 
correlation between the trimer and tetramer binding energies
in Fig.~\ref{fig:scale4}.
%%%%%%%%%%%%%%%%%%%%%%%%%%%%%%%%%%%%%%%%%%%%%%%%%%%%%%%%%%%%%%%%%%%%%%%
%%%%%%%%%%%%%%%%%%%%%%%%%%%%%%%%%%%%%%%%%%%%%%%%%%%%%%%%%%%%%%%%%%%%%%%%
The left panel shows the correlation between the ground state
energies while the right panel shows the correlation between the 
excited state energies. The data points are calculations using various 
methods and $^4$He potentials (see Ref.~\cite{Blume:2000} and references
therein). We conclude that universality is well satisfied in the three-
and four-body systems of $^4$He atoms.

The existence of these scaling functions is a universal feature 
of systems with large scattering length and is independent of the 
details of the short-distance physics. Similar correlations
between few-body observables appear for example also in nuclear
systems. The correlation between the triton and $\alpha$ particle
binding energies (the Tjon line) can be explained using the same
effective theory \cite{Platter:tjon}.

\section{$N$-BOSON DROPLETS IN 2D}

In this section, we consider the universal properties of
weakly interacting bosons with large scattering length
(or equivalently a shallow dimer state)
in two spatial dimensions (2D) \cite{Hammer:2004as}.  
In particular, we consider self-bound droplets of $N(\gg1)$
bosons interacting weakly via an {\em attractive}, short-ranged 
pair potential. Our analysis relies strongly on the
property of asymptotic freedom of 2D bosons with an attractive
interaction.

In 2D, any attractive potential has at least one bound state.  For the
potential $-g\delta^2(r)$ with small $g$, there is exactly one bound
state with an exponentially small binding energy,
$
  B_2 \sim \Lambda^2\exp\left( - {4\pi}/g\right),
$
where $\Lambda$ is the ultraviolet momentum cutoff (which is the
inverse of the range of the potential).
Asymptotic freedom provides an elegant way to understand
this result.  In 2D nonrelativistic theory, the four-boson
interaction $g(\psi^\dagger\psi)^2$ is marginal.  The coupling runs
logarithmically with the length scale $R$, and the running can be
found by performing the standard renormalization group (RG) procedure.
For $g>0$, the coupling grows in the
infrared, in a manner similar to the QCD
coupling.  The dependence of the coupling
on the length scale $R$ is given by
\begin{equation}\label{running}
  g(R) = \left[\frac{1}{g} - \frac{1}{4\pi}\ln 
        (\Lambda^2 R^2)\right]^{-1}\,,
\end{equation}
so the coupling becomes large when $R$ is comparable to the size of
the two body bound state $B_2^{-1/2}$. This is in essence the
phenomenon of dimensional transmutation: a dynamical scale is
generated by the coupling constant and the cutoff scale.
It is natural, then, that $B_2$ is the only physical energy scale in
the problem: the binding energy of three-particle, four-particle,
etc. bound states are proportional to $B_2$.  In contrast to 
three spatial dimensions, there is no Efimov effect (or Thomas collapse)
and no three-body parameter is required. The
$N$-particle binding energy $B_N$, however, 
can be very different from $B_2$ if
$N$ is parametrically large.  We use the
variational method to calculate the size of the bound state.  For a
cluster of a large number of bosons, one can apply classical
field theory.  We thus have to minimize the expectation
value of the Hamiltonian with respect to all field configurations $\psi(r)$
satisfying the constraint $N = \int\!d^2r\, \psi^\dagger\psi\,$.
This is equivalent to a Hartree calculation with the running
coupling constant $g(R)$ instead of the bare one. In the limit
of a large number $N$ of particles in the droplet, some exact 
predictions can be obtained \cite{Hammer:2004as}.

The system possesses surprising universal
properties.  Namely, if one denotes the size of the $N$-body droplet
as $R_N$, then at large $N$ and in the limit of zero range of the
interaction potential:
\begin{equation}
  \label{RNratio}
  {R_{N+1}}/{R_N} \approx 0.3417,\qquad N\gg 1\,.
\end{equation}
The size of the bound state decreases exponentially with $N$: adding a
boson into an existing $N$-boson droplet reduces the size of the
droplet by almost a factor of three.  Correspondingly, the binding energy of
$N$ bosons $B_N$ increases exponentially with $N$:
\begin{equation}
  \label{BNratio}
  {B_{N+1}}/{B_N} \approx 8.567, \qquad N\gg 1\,.
\end{equation}
This implies that the energy required to remove
one particle from a $N$-body bound state (the analog of the
nucleon separation energy for nuclei) is about 88\% of the
total binding energy.  This is in contrast to most other physical
systems, where separating one particle costs much less energy than the
total binding energy, provided the number of particles in the bound
state is large. The $1/N$-corrections to Eqs.~(\ref{RNratio}, \ref{BNratio})
are calculable.

For the universal predictions (\ref{RNratio}, \ref{BNratio}) to apply
in realistic systems with finite-range interactions,
the $N$-body bound states need to be sufficiently shallow and hence 
have a size $R_N$ large compared to the natural low-energy length scale $l$.
Depending on the physical system,
$l$ can be the van der Waals length $l_{vdW}$, the range of the 
potential or some other scale. As a consequence,
Eqs.~(\ref{RNratio}, \ref{BNratio}) are valid in such systems
for $N$ large, but below a critical value,
\begin{equation}
  1 \ll N \ll N_{\rm crit} \approx 0.931 \ln({R_2}/{l})
  + {\cal O}(1)\,.
\label{eq:break}
\end{equation}
At $N=N_{\rm crit}$ the size of the droplet is comparable to $l$
and universality is lost.  If there is an exponentially
large separation between $R_2$ 
and $l$, then $N_{\rm crit}$ is much larger than one and 
the condition (\ref{eq:break}) can be satisfied.

We can compare our prediction with exact few-body calculations
for $N=3,4$. While the $1/N$-corrections are expected to be
relatively large in this case, we can estimate how the universal result 
(\ref{BNratio}) is approached.
The three-body system for a zero-range potential in 2D has exactly two
bound states: the ground state with
$B_3^{(0)}=16.522688(1)\, B_2$ and one excited state with
$B_3^{(1)}=1.2704091(1)\, B_2$ \cite{Bruch,Nielsen,Hammer:2004as}.
Similarly, the four-body system for a zero-range potential in 2D 
has two bound states: the ground 
state with $B_4^{(0)}=197.3(1)B_2$ and one excited state with 
$B_4^{(1)}=25.5(1)B_2$ \cite{Platter:2004ns}.
The prediction (\ref{BNratio}) applies to the ground state energies
$B_3^{(0)}$ and $B_4^{(0)}$.
The ratio $B_3^{(0)}/B_2\approx 16.5$ is almost twice as large as the 
asymptotic value~(\ref{BNratio}), while the ratio $B_3^{(0)}/B_4^{(0)}
\approx 11.9$ is already considerably closer.
These deviations are expected for such small 
values of $N$. Note, however, that the ratio of
the root mean square radii of the two- and three-body wave functions
is $0.306$~\cite{Nielsen}, close to the asymptotic
value~(\ref{RNratio}). 

It would be interesting to test the universal 
predictions (\ref{RNratio}, \ref{BNratio}) both theoretically and 
experimentally for $N>4$. On the theoretical side, Monte Carlo 
techniques appear to be a promising avenue.
Furthermore, the experimental realizablity of self-bound 
2D boson systems with weak interactions should be investigated.

\section{SUMMARY \& OUTLOOK}

We have discussed the EFT for few-body systems with
short-range interactions and large scattering length $a$. The main focus
was on applications to cold atoms.

The renormalization of the three-body system with large $a$
in three spatial dimensions
requires a one-parameter three-body force governed by a limit
cycle already at leading order in the expansion in $l/|a|$.
As a consequence, two parameters are required to specify a three-body system:
the scattering length $a$ (or the dimer
binding energy $B_2$) and the three-body parameter $\Lambda_*$.
However, once these two parameters are given the properties of the 
three- and four-body systems are fully 
determined at leading order in $l/|a|$.

The large scattering length leads to universal properties
independent of the short-distance dynamics. In particular, we have
discussed universal expressions for three-body observables and 
universal scaling functions relating different few-body observables.

In two spatial dimensions, the three-body parameter $\Lambda_*$ does
not enter at leading order in the expansion in $l/|a|$ and
$N$-body binding energies only depend on $B_2$. The asymptotic freedom
of non-relativistic bosons with attractive interactions in 2D leads
to some remarkable universal properties of $N$-body droplets.

Future challenges include the application to Halo systems in nuclear physics
\cite{Bertulani:2002sz},
and the possibility of coexisting condensates of atoms, dimers, and trimers
in atomic physics \cite{Braaten:2002er}. 
The three-body effects discussed here will also
become relevant in Fermi systems with three or more spin states.

%%%%%%%%%%%%%%%%%%%%%%%%%%%%%%%%%%%%%%%%%%%%%%%%
%% BACKMATTER
%%%%%%%%%%%%%%%%%%%%%%%%%%%%%%%%%%%%%%%%%%%%%%%%

\begin{theacknowledgments}
This work was done in collaboration with E.~Braaten, U.-G.~Mei\ss ner,
L.~Platter, and D.T.~Son. This research was supported by the US Department of
Energy under grant DE-FG02-00ER41132.
\end{theacknowledgments}

%% If the bibliography is produced without BibTeX comment out the

\bibliographystyle{aipproc}   % if natbib is available

\end{document}